\documentclass[showpacs]{revtex4}

\usepackage{graphicx}

\newcommand{\be}[1]{\begin{equation}\label{#1}}
\newcommand{\ee}{\end{equation}}

\begin{document}

\title{Stylized facts from a threshold-based
    heterogeneous agent model}

\author{R. Cross}\affiliation{Department of Economics,
University of Strathclyde,
Sir William Duncan  Building, 130 Rottenrow
Glasgow G4 0GE, Scotland, UK}

\author{M. Grinfeld} \affiliation{Department of Mathematics,
University of Strathclyde, Livingstone Tower, 26 Richmond Street,
Glasgow G1 1XH, Scotland, UK}

\author{H. Lamba} \affiliation{Department of Mathematical Sciences,
George Mason University, 4400 University Drive,
Fairfax, VA 22030 USA}

\author{T. Seaman} \affiliation{School of Computational Sciences,
George Mason University, 4400 University Drive,
Fairfax, VA 22030 USA}

\date{\today}

\begin{abstract}
A class of heterogeneous agent models is investigated where investors
switch trading position whenever  their motivation to do so
exceeds some critical threshold. These motivations can be
psychological in nature or reflect behaviour suggested by
the efficient market hypothesis (EMH).
By introducing different propensities into a baseline
model that displays EMH
behaviour, one can attempt to isolate their effects upon the market dynamics. 

The simulation results indicate that the introduction of a herding propensity
results in excess kurtosis and power-law decay consistent with those
observed in actual return distributions, but not in significant
long-term volatility
correlations. Possible alternatives  for introducing such long-term
volatility correlations are then identified and discussed.
\end{abstract}



\maketitle

\section{\label{intro}Introduction}  

It is hard to overestimate the impact that the concept of efficient
markets has had on economic and political thinking. 
The underlying efficient market hypothesis (EMH) \cite{f70} has enormous 
philosophical and mathematical appeal but is perhaps
best thought of as a Platonic ideal.
The strong form of the hypothesis is that investors 
have access to all relevant information, and that this is fully
reflected by the current market price.
The random arrival of new
(independent and identically Gaussian-distributed) information causes traders'
expectations to change. 
This is then translated into a Brownian motion in, and a Gaussian
distribution of, (log) price returns. There are variations upon the
above reasoning, for example, invoking arbitrageurs or `informed' investors who
quickly exploit any inefficiencies due to `noise' traders or
`uninformed' investors but the pricing outcome is the same.   
One of the refutable implications 
of the EMH is the Gaussian distribution of returns. Actual distributions 
however are sufficiently non-Gaussian so as to require better explanations
and mathematical models than provided by the EMH \cite{ms00,c01}.

Two types of assumptions underlie the EMH.  Firstly, there are
assumptions about the nature of the information entering the system
(for example, its stationarity and lack of correlations), 
the dissemination of this data amongst the market participants, and their ability to
evaluate and react to it. Given the enormous increase in information
processing speeds, and the 
rise of instantaneous mass global communication,
it is not implausible to suppose that some EMH violations of this type
have become less important  over recent decades.

The second set of assumptions concerns the
rationality and motivations of the agents themselves, be they
individuals or financial institutions. As regards individuals,
recent work by psychologists
and experimental economists has suggested that deviations from
expected utility maximisation are widespread, even when `smart' people are
playing `simple' games.
Furthermore, there are structural and institutional features
that can undermine the EMH. Examples include
compensation/evaluation/bonus criteria, tax laws, accounting rules,
conflicts of 
interest within a financial organization and  moral hazards problems.

With so many plausible EMH violations (and the impossibility of
performing controlled experiments with real markets),
it is extremely difficult to draw conclusions
regarding the chain of cause and effect from statistical analyses
alone. However these analyses have identified
a set of `stylized facts' that appear to be
prevalent across  asset classes independent of trading rules,
geography or culture. These include the lack of linear correlations in price
returns over all but the shortest timescales, excess kurtosis
(fat-tails) in the price return distribution, volatility clustering
and heteroskedasticity. Some finer details have also been revealed, most notably the existence of
power-law scalings and estimates of the exponents.

The class of models
presented here (see also \cite{cgls05,ls06,l06}) is an attempt to provide a framework
within which to study systematically the effects of various, simple,
EMH violations. The hope is that the insights gained will result in
both a greater theoretical understanding of the operation of markets
and in better simulation tools for market practitioners. 
The modeling process we advocate is based upon the idea of thresholds. At each
point in time, agents are comfortable with their current position
(either long or short on the market). However, they are subject to one or
more `tensions' which cause a switch in position whenever the
corresponding threshold is violated. The use of the word `tension'
does not necessarily imply that the response is emotional or
psychological in
nature (although it may be) --- the agent may have buy/sell price triggers
in place based upon analytical research, in which case the tension
level merely reflects the distance from the current price to the
closest threshold. 

These models, together with a related approach that can be applied to
Minority Games \cite{cglp04},  have been introduced
elsewhere \cite{cgls05,ls06}  and the reader is directed to them for
further details. The main contributions of this paper are to
more thoroughly consider the modeling of volatility clustering 
and examine the relative performances of the heterogeneous agents. 
The paper is organized  as follows.
In Section~\ref{sec2} we introduce a minimal, baseline,  model in which the market price remains
identical to a market operating under the EMH. By including
additional tensions one can then  observe the corresponding changes in the
market statistics. This is performed  in Section~\ref{sec3},
where a herding propensity is included, resulting in fat-tails and
excess kurtosis, but no long-term volatility correlations. 
In Section~\ref{sec4} we discuss different possibilities for
generating volatility clustering in the form of slowly-decaying correlations.
Finally, in Section~\ref{pal} the relative performance of agents with
differing herding propensities is investigated.

\section{\label{sec2}A threshold model with EMH price returns}

The system evolves in discrete timesteps of length $h$ (which  will be
 chosen to correspond to one trading day for the simulations in this paper).
 There are $M$ agents, all of equal size, who can be either  
long or short in the market  over
the $n^{\rm th}$ time interval. The market price at the end of the
$n^{\rm th}$ time interval is $p(n)$. 
For simplicity $p(0)= 1$ and  we assume that the system is
 drift-free so that, in reality, $p(n)$ corresponds to, say, the price
 corrected for the risk-free interest rate plus equity-risk premium or
 the expected rate of return.
The position of the $ i^{th} $ investor over the $ n^{th} $ time
interval is represented by
  $s_i(n) = \pm 1 $ ($+1$ long, $-1$ short), and the sentiment of
  the market 
by the average of the states of all of the $M$ investors
\begin{equation} \sigma (n) = \frac{1}{M} \sum_{i=1}^M s_i(n).\label{sigma} \end{equation}
The change in market sentiment from the previous time interval 
is defined  by $\Delta \sigma (n) = \sigma (n) - \sigma(n-1)$.

Before defining the model we make the following important
point. We are not attempting to simulate directly all of the market
participants, just those whose trading strategies
 are most significant over the timescale of interest. Thus we 
start by hypothesizing the existence of some underlying EMH market and
change as little as possible. In
particular we shall assume that arbitrageurs and traders  
exist who act to interpret the incoming information
stream and induce the corresponding price changes over timescales
$\ll h$. Other market details,
such as the way in which orders are placed and executed,
remain unspecified but constant. 

 We shall also assume a simple linear relationship between changes in
 the sentiment $\Delta \sigma$ and the excess pricing pressure it induces. 
This leads us to the following geometric pricing formula
  \begin{equation}
p(n+1)=p(n) \exp\left(\sqrt{h}\eta(n) - h/2 + \kappa \Delta\sigma(n)\right)\label{price0}
\end{equation}
where $\sqrt{h}\eta(n) \sim {\cal N}(0,h)$ represents the exogenous
information stream. The 
parameter $\kappa$
reflects the relative effects on price of internally generated dynamics
as opposed to the information. Finally, the term $-h/2$ is the drift
correction required by It\^{o} calculus to ensure that, for $\kappa =
0$, the price $p(t)$ is a martingale. It can be safely omitted from
the model but we choose to include it here for completeness.

In order to close the model we must now specify how the states of the
individual agents are determined, i.e. how the $i^{\rm th}$ agent
decides when to switch. 
This is achieved by introducing an `inaction' pressure. Every time the
agent switches position a pair of threshold prices on either side of
the current price is generated. When the current market price crosses one of these  
threshold values the agent switches once again, a new pair of
thresholds is generated and the process repeats (more generally, the
thresholds can be updated continuously rather 
than only when the agent switches but this appears to make little
difference to the behaviour of the model). An appealing feature of
the inaction
pressure is that it is capable of multiple interpretations
 --- at the `rational' end of
the spectrum, the price interval defined by the
thresholds corresponds to  an
investment strategy based upon the market analysis and future
expectations of that agent.
 Other effects that can also
be reproduced, are: the psychological factors behind the desire to cut
losses or take profits; transaction costs 
and the resulting hysteresis effects; the irrational need for
agents to do something or the (less ir)rational need to be seen to be doing
something (in the case of active-fund managers, perhaps). Further 
details can be found in \cite{cgls05}.

To define the model precisely, 
let $P_i$ be the price at which the $i^{th}$ investor last switched
  positions and let $H_i > 0$ be a value, chosen
  randomly at each switching 
from the uniform distribution on the  interval $[H_{\rm L}, H_{\rm U}]$.
Then, as long as the current price $p(n)$ stays within the interval 
  $ [P_i / (1+H_i), \  P_i (1+H_i) ],$ the investor maintains her position,
but if the current price $p(n)$ leaves this interval, the investor
switches. The choice of a uniform distribution is made purely on grounds of
simplicity --- the model appears to be extremely robust and, in the absence of
other information, there is nothing to be gained by making the model more
complicated than necessary.

The behaviour of the above model is reasonably straightforward. Provided that
$M$ is sufficiently large ($M=100$ appears to be sufficient),  and
that the
initial agent states are sufficiently mixed with $\sigma(0) \approx 0$,
 sentiment will remain close to $0$ and the price remains
close to its fundamental EMH value. This is because there is no
coupling between agents and their switches in position
cancel without affecting the sentiment \cite{m03}. Thus we have a model that
is very close, both philosophically and in appearance, to that posited by the EMH --- the
price follows a geometric Brownian motion and, if one interprets the
inaction pressure in the `rational' way described above,  trading is induced by
the differing expectations of agents. We hesitate to
describe the model as efficient since the volume of trading is
determined solely  by the interval $[H_{\rm L}, H_{\rm U}]$. This
implies that excess trading may occur which is inefficient in the presence
of transaction costs. However such excess trading is another well-documented
feature of actual financial markets \cite{s00}.

\section{\label{sec3}Incorporating a herding pressure}  

There are other pressures affecting investors which, when included in the model, will
not not necessarily cancel out, most likely due to some form of global coupling.  
The simplest, and arguably the single most important, example of such a pressure
is the `herding tendency' --- while an
individual/organization  is holding a minority opinion/position
they may feel an increasing pressure to conform that eventually becomes
unbearable (unless enough  of
the agents with majority positions switch first),  at
which point they will switch to join the majority. Clearly 
different agents will have different tolerance levels that are, to some
extent, a reflection of their personality or trading philosophy (such
as `momentum traders' and `contrarian
investors'). Although it is tempting to describe
such herding behaviour as irrational, or `boundedly-rational' in the sense of
Simon \cite{s55,s97}, this may not be a fair characterization in all
cases. Some agents  may lose their job/investment capital if they
significantly underperformed the average market or benchmark 
return for even a few quarters in a
row --- such agents are exhibiting behaviour that is no more
irrational than animals herding when surrounded by predators \cite{c04}.

We incorporate the herding tendency as follows.
At time $n$, the  herding pressure felt by agent $i$ is denoted by $c_i(n)$. 
This level is changed to $c_i(n+1) = c_i(n) + h | \sigma(n)
|$ (i.e. is  increased by an 
  amount proportional to the length of the time interval  and the 
   severity of the inconsistency) whenever $ s_i(n) \sigma(n) < 0. $ 
Otherwise, the agent's herding pressure remains unchanged and $c_i(n+1) = c_i(n).$
As soon as  $ c_i(n) $ exceeds her (constant)
threshold $C_i,$ the investor switches market position and $c_i$ is
reset to zero.
Additionally we suppose that whenever a switch occurs, both the
inaction and herding pressures are set to zero (although the model
appears to be very robust with respect to such changes in the interactions
between the tensions \cite{cgls05,ls06}).

We now choose some realistic parameters and present some numerical
results. A daily variance in price returns of 0.6--0.7\%
suggests a value for $h$ of 0.00004. The number of participants
$M=100$ and it is worth noting that the model's characteristics are independent of
$M$ --- this is an important property not always shared  by other
heterogeneous agent models.
The simulation is run for 10000 timesteps which corresponds to
approximately 40 years of trading.

Once $h$ has been fixed, we suppose that the $C_i$ are
chosen from the uniform distribution on $[0.001,0.004],$ as this leads
to herd-induced switching on the timescale of weeks and months for those
agents in the minority. The price ranges for the inaction tension are
chosen randomly after every switching from the uniform distribution on
the interval $10\% - 30\%$, i.e. $[H_L, H_U] = [0.1,0.3]$. Day-traders
would of course have much smaller values but our choice of $h$ means
that we cannot attempt to model directly changes occurring over such
short timescales. Finally,  simulations
using the above parameters suggest that a value of $\kappa=0.2$
results in prices  that are strongly correlated with the information
stream  
but which differ significantly during periods of
extreme market sentiment.

\begin{figure}
  \includegraphics[width=3in]{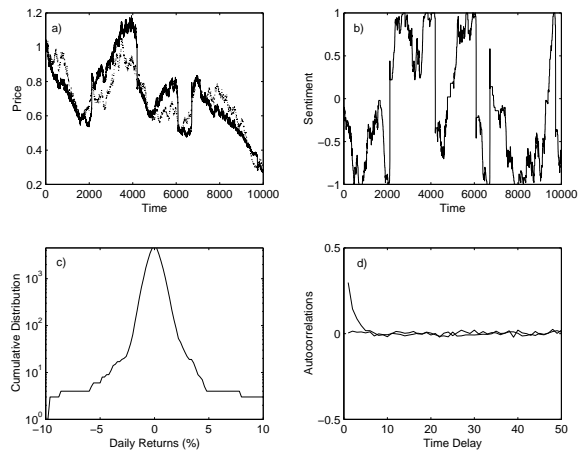}
\caption{\label{fig1} Results of a simulation over 10000
  timesteps. See the text for details.}
\end{figure}

Figure~\ref{fig1} shows the output of a typical run. Figure \ref{fig1}a) plots the
price $p(t)$ against the `fundamental' price obtained by setting $\kappa =
0$ (which decouples the price from the agent dynamics and generates a
pure geometric Gaussian price stream). 
It should be noted that the agents typically switch every few weeks or
months and that the vast majority of trades are due to the inaction thresholds
being violated. However the sentiment $\sigma$, as can be seen in
Figure~\ref{fig1}b), changes more slowly and can remain bullish or bearish for
several years. Figure~\ref{fig1}c) plots the daily log price returns.
Fat-tails displaying power-law behaviour with exponents in the range
$[2.8,3.2]$ are observed \cite{ls06} (together with kurtosis values in the
approximate range $[10,50]$). Finally Figure~\ref{fig1}d) plots the
autocorrelation functions of the price returns and absolute price
returns, a measure of volatility. The price returns show no evidence of linear
correlations even with a lag of one day. The volatility correlations
however die away after approximately 5 days or so. This lack of
long-term volatility correlations or memory is the subject of the next section.

To recap, the introduction of
herding does indeed generate fat-tails with decay rates that fit
values extracted from actual market data. Further details,
together with a `computational experiment' that shows how to generate
second-order effects such as observed asymmetries in the price return
data with respect to positive and negative price moves can be found in \cite{ls06}.

\section{\label{sec4}Simulating clustered volatility}  

Market models must be able to approximate 
the statistical properties of the market
volatility which we  define as the absolute log-price 
return $\left|\strut \log\frac{p(n+1)}{(p(n)}\right|$. 
However the causes of volatility clustering and long-memory are still poorly
  understood and there are several plausible mechanisms, all
  of which may play a significant role. There have been numerous
  studies investigating the relationship between volatility and other
  market variables, such as trading volume, but the question is still
  far from being resolved.

One possibility is that the clustering is due to
non-stationarity and/or long-time correlations in the data stream. This is
certainly plausible --- geopolitical events and changes in economic
conditions are rarely revealed by a single pulse of information entering the market,
but rather unfold over a period of time. For the models of Section~\ref{sec2} and \ref{sec3} these
effects 
could be incorporated  by replacing  $\eta(n)$ with time series derived from fractional
Brownian processes, stochastic volatility models,  or
GARCH-type processes (although one must be careful 
to ensure that no correlations are introduced into the returns
themselves \cite{bbm96}).  However, certainly within the
context of heterogeneous agent models (HAMs), these possibilities tend
to be ignored,  perhaps because it is more interesting to develop market
`black boxes' where all the non-Gaussian effects are generated
internally.
It is also possible to generate volatility clustering within HAMs via 
inductive learning and evolutionary strategies. To include
such effects into our threshold models is certainly achievable (by choosing the
inaction thresholds $H_i$ to reflect the agents' current strategy) but
the resulting models are extremely complex and will not be considered here.

In the majority of HAMs that
display clustered volatility, the underlying mechanism appears to be
the ability of agents to switch between different `fundamentalist'
 and `chartist'  strategies (for example, the Lux-Marchesi model \cite{lm00}). 
Fundamentalist traders are betting that the price
 will quickly revert to some underlying rational price while the
 chartists believe that the recent price-trend will continue. Bubbles and
 anti-bubbles occur whenever the proportion of chartists exceeds some
 critical value. In our threshold models the agents are all of the same
 qualitative type so this switching between strategies cannot
 occur. However, the $M$ agents being explicitly simulated do not
 constitute the entire market since short-term noise traders are excluded.   
We now hypothesize that the number and activity-level of these traders is
not constant in time but instead depends upon market conditions. The
simplest scenario is that their effect upon the market is a function
of overall sentiment. There is some evidence to support this
correlation between volatility and (both bullish and bearish) sentiment from closed-end investment
funds \cite{b99} (together with strong indications that the increases in
volatility during times of extreme market sentiment were indeed due to
noise traders rather than excess trading by fundamentalist investors).

\begin{figure}
  \includegraphics[width=3in]{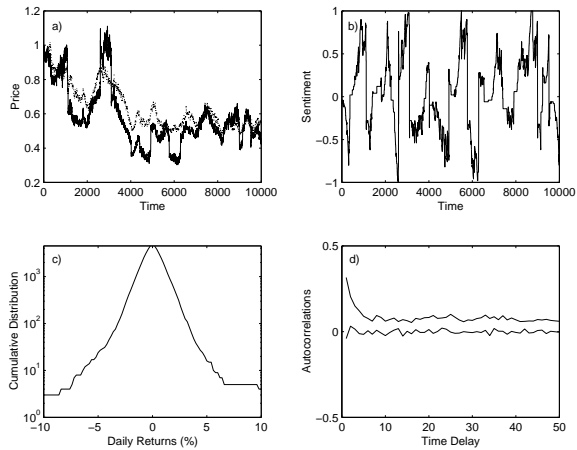}
\caption{\label{fig2}The same data are plotted as in
    Figure~1 with but using the pricing formula (\ref{price1}). }
\end{figure}

Thus we replace the pricing formula (\ref{price0}) with
  \begin{equation}
p(n+1)=p(n) \exp\left(\left(\sqrt{h}\eta(n) - h/2\right)
f(\sigma) + \kappa \Delta\sigma(n)\right)\label{price1}
\end{equation}
and assume a simple linear dependence of $f$ upon $|\sigma|$,
i.e. $f(\sigma) = 1 + \alpha|\sigma| $ (setting $\alpha =  0$ reverts to
the model of Section~\ref{sec3}).
A value of $\alpha  = 2$ (keeping all other parameters unchanged from
Section~\ref{sec3}) does indeed add volatility clustering as can
be seen in Figure~\ref{fig2}. The rate of decay of the
volatility autocorrelation function is an approximate power-law with exponent in the range 0.3--0.5.

It should be noted that these threshold models are 
non-Markovian since the agents' tension levels are highly dependent
upon the past behaviour of the system. This memory effect seems to be
fundamental to the formation and collapse of the extended periods
of mis-pricing that occur (and the corresponding fat-tails). 
However, the long-time volatility correlation introduced by
(\ref{price1}) is not due to memory-effects. Rather, the
price volatility due to external information now depends, via the
function $f(\sigma)$ in (\ref{price1}), upon a 
slowly-changing system variable, namely the sentiment $\sigma$, and
inherits its slow autocorrelation decay.  

\section{\label{pal}Profitability of traders} 

Finally we perform an interesting numerical experiment.  Note that the
agents' inaction thresholds change at every switching (to reflect
updated future expectations) but their herding thresholds do not. This
is because we consider the latter to be a measure of each agent's
trading philosophy or personality and so more likely to remain constant
over time. This raises the question of whether there is an observable
difference in the relative performance between  agents whose  
threshold values $C_i$ lie within the range $[0.001,0.004]$ used in the
simulations. Such a difference would suggest, within this modeling
framework, the possibility of elementary, but effective, inductive
learning strategies that simply consist of agents `training' themselves to
change their herding propensity.

To answer this question we keep track of
the agents' profit or loss at each transaction during the simulation (note that the agents' 
financial performance does not affect their behaviour, 
although the reproduction of more realistic
psychological pressures would probably include factors such as
these). The agents are always assumed to hold $\pm 1$ units of the
underlying asset and an inexhaustible cash supply to fund the transactions.
The performance over the first 1000 timesteps is ignored to exclude
transient effects caused by the externally imposed initial conditions.

\begin{figure}
  \includegraphics[width=2.5in]{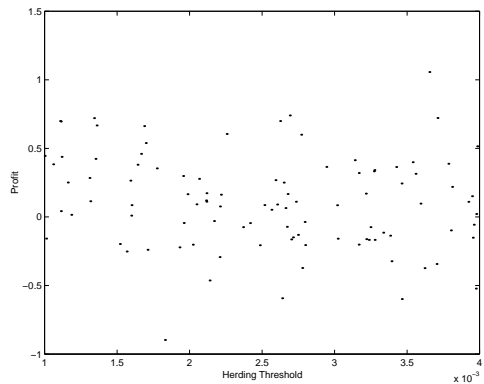}
\caption{\label{fig3}The profit/loss of each agent
    plotted against their herding threshold. No significant
    correlation is observed.}
\end{figure}

The performance of the agents is displayed in Figure~\ref{fig3} where
the overall profit/loss is plotted against that agent's herding
threshold $C_i$.
There is no significant correlation between profits and herding
propensity. And of course if transaction costs are taken into account
then agents with lower thresholds would actually perform relatively worse.

\section{\label{conc}Concluding Remarks} 

The class of threshold  HAMs studied here can incorporate enough psychology to describe
realistic market behaviour. They are, however, difficult to
analyse. But since all the 
coupling is global, a mean-field approach is possible. The resulting
objects are stochastic difference equations coupled to deterministic ones;
see \cite{cgl06} for an initial study of such a model, which, surprisingly, manages
to reproduce some degree of volatility clustering without an explicit
mechanism such as in (3). In future, we aim to use methods of discrete
random dynamical systems \cite{df} in order to elucidate, inter alia, the reasons
for the appearance of power laws in the system.


\begin{thebibliography}{17}
\expandafter\ifx\csname natexlab\endcsname\relax\def\natexlab#1{#1}\fi
\expandafter\ifx\csname bibnamefont\endcsname\relax
  \def\bibnamefont#1{#1}\fi
\expandafter\ifx\csname bibfnamefont\endcsname\relax
  \def\bibfnamefont#1{#1}\fi
\expandafter\ifx\csname citenamefont\endcsname\relax
  \def\citenamefont#1{#1}\fi
\expandafter\ifx\csname url\endcsname\relax
  \def\url#1{\texttt{#1}}\fi
\expandafter\ifx\csname urlprefix\endcsname\relax\def\urlprefix{URL }\fi
\providecommand{\bibinfo}[2]{#2}
\providecommand{\eprint}[2][]{\url{#2}}

\bibitem[{\citenamefont{Fama}(1970)}]{f70}
\bibinfo{author}{\bibfnamefont{E.}~\bibnamefont{Fama}}, \bibinfo{journal}{J.
  Finance} \textbf{\bibinfo{volume}{25}}, \bibinfo{pages}{383}
  (\bibinfo{year}{1970}).

\bibitem[{\citenamefont{Mantegna and Stanley}(2000)}]{ms00}
\bibinfo{author}{\bibfnamefont{R.}~\bibnamefont{Mantegna}} \bibnamefont{and}
  \bibinfo{author}{\bibfnamefont{H.}~\bibnamefont{Stanley}},
  \emph{\bibinfo{title}{An Introduction to Econophysics}}
  (\bibinfo{publisher}{CUP}, \bibinfo{year}{2000}).

\bibitem[{\citenamefont{Cont}(2001)}]{c01}
\bibinfo{author}{\bibfnamefont{R.}~\bibnamefont{Cont}},
  \bibinfo{journal}{Quantitive Finance} \textbf{\bibinfo{volume}{1}},
  \bibinfo{pages}{223} (\bibinfo{year}{2001}).

\bibitem[{\citenamefont{Cross et~al.}(2005{\natexlab{a}})\citenamefont{Cross,
  Grinfeld, Lamba, and Seaman}}]{cgls05}
\bibinfo{author}{\bibfnamefont{R.}~\bibnamefont{Cross}},
  \bibinfo{author}{\bibfnamefont{M.}~\bibnamefont{Grinfeld}},
  \bibinfo{author}{\bibfnamefont{H.}~\bibnamefont{Lamba}}, \bibnamefont{and}
  \bibinfo{author}{\bibfnamefont{T.}~\bibnamefont{Seaman}},
  \bibinfo{journal}{Phys. A} \textbf{\bibinfo{volume}{354}},
  \bibinfo{pages}{463} (\bibinfo{year}{2005}{\natexlab{a}}).

\bibitem[{\citenamefont{Lamba and Seaman}()}]{ls06}
\bibinfo{author}{\bibfnamefont{H.}~\bibnamefont{Lamba}} \bibnamefont{and}
  \bibinfo{author}{\bibfnamefont{T.}~\bibnamefont{Seaman}},
  \bibinfo{note}{preprint, Econophysics forum}.

\bibitem[{\citenamefont{Le{B}aron}(2006)}]{l06}
\bibinfo{author}{\bibfnamefont{B.}~\bibnamefont{Le{B}aron}}, in
  \emph{\bibinfo{booktitle}{Post-{W}alrasian Economics}}, edited by
  \bibinfo{editor}{\bibfnamefont{D.}~\bibnamefont{Colander}}
  (\bibinfo{publisher}{CUP, New York}, \bibinfo{year}{2006}).

\bibitem[{\citenamefont{Cross et~al.}(2005{\natexlab{b}})\citenamefont{Cross,
  Grinfeld, Lamba, and Pittock}}]{cglp04}
\bibinfo{author}{\bibfnamefont{R.}~\bibnamefont{Cross}},
  \bibinfo{author}{\bibfnamefont{M.}~\bibnamefont{Grinfeld}},
  \bibinfo{author}{\bibfnamefont{H.}~\bibnamefont{Lamba}}, \bibnamefont{and}
  \bibinfo{author}{\bibfnamefont{A.}~\bibnamefont{Pittock}}, in
  \emph{\bibinfo{booktitle}{Relaxation Oscillations and Hysteresis}}, edited by
  \bibinfo{editor}{\bibfnamefont{M.}~\bibnamefont{Mortell}},
  \bibinfo{editor}{\bibfnamefont{R.~O.} \bibnamefont{Jr.}},
  \bibinfo{editor}{\bibfnamefont{A.}~\bibnamefont{Pokrovskii}},
  \bibnamefont{and} \bibinfo{editor}{\bibfnamefont{V.}~\bibnamefont{Sobolev}}
  (\bibinfo{publisher}{SIAM}, \bibinfo{year}{2005}{\natexlab{b}}), pp.
  \bibinfo{pages}{61--72}.

\bibitem[{\citenamefont{Malkiel}(2003)}]{m03}
\bibinfo{author}{\bibfnamefont{B.}~\bibnamefont{Malkiel}},
  \bibinfo{journal}{Journal of Economic Perspectives}
  \textbf{\bibinfo{volume}{17}}, \bibinfo{pages}{59} (\bibinfo{year}{2003}).

\bibitem[{\citenamefont{Schleifer}(2000)}]{s00}
\bibinfo{author}{\bibfnamefont{A.}~\bibnamefont{Schleifer}},
  \emph{\bibinfo{title}{Inefficient Markets}}, Clarendon Lectures in Economics
  (\bibinfo{publisher}{OUP}, \bibinfo{year}{2000}).

\bibitem[{\citenamefont{Simon}(1955)}]{s55}
\bibinfo{author}{\bibfnamefont{H.}~\bibnamefont{Simon}},
  \bibinfo{journal}{Quart. J. Econ.} \textbf{\bibinfo{volume}{69}},
  \bibinfo{pages}{99} (\bibinfo{year}{1955}).

\bibitem[{\citenamefont{Simon}(1997)}]{s97}
\bibinfo{author}{\bibfnamefont{H.}~\bibnamefont{Simon}},
  \emph{\bibinfo{title}{Models of Bounded Rationality}}
  (\bibinfo{publisher}{MIT Press}, \bibinfo{year}{1997}).

\bibitem[{\citenamefont{Chamley}(2004)}]{c04}
\bibinfo{author}{\bibfnamefont{C.}~\bibnamefont{Chamley}},
  \emph{\bibinfo{title}{Rational Herds}} (\bibinfo{publisher}{CUP},
  \bibinfo{year}{2004}).

\bibitem[{\citenamefont{Baillie et~al.}(1996)\citenamefont{Baillie, Bollerslev,
  and Mikkelsen}}]{bbm96}
\bibinfo{author}{\bibfnamefont{R.}~\bibnamefont{Baillie}},
  \bibinfo{author}{\bibfnamefont{T.}~\bibnamefont{Bollerslev}},
  \bibnamefont{and}
  \bibinfo{author}{\bibfnamefont{H.}~\bibnamefont{Mikkelsen}},
  \bibinfo{journal}{J. Econometrics} pp. \bibinfo{pages}{3--30}
  (\bibinfo{year}{1996}).

\bibitem[{\citenamefont{Lux and Marchesi}(2000)}]{lm00}
\bibinfo{author}{\bibfnamefont{T.}~\bibnamefont{Lux}} \bibnamefont{and}
  \bibinfo{author}{\bibfnamefont{M.}~\bibnamefont{Marchesi}},
  \bibinfo{journal}{Int. J. Theor. Appl. Finance} \textbf{\bibinfo{volume}{3}},
  \bibinfo{pages}{675} (\bibinfo{year}{2000}).

\bibitem[{\citenamefont{Brown}(1999)}]{b99}
\bibinfo{author}{\bibfnamefont{G.}~\bibnamefont{Brown}},
  \bibinfo{journal}{Financial Analysts Journal} pp. \bibinfo{pages}{82--90}
  (\bibinfo{year}{1999}).

\bibitem[{\citenamefont{Cross et~al.}()\citenamefont{Cross, Grinfeld, and
  Lamba}}]{cgl06}
\bibinfo{author}{\bibfnamefont{R.}~\bibnamefont{Cross}},
  \bibinfo{author}{\bibfnamefont{M.}~\bibnamefont{Grinfeld}}, \bibnamefont{and}
  \bibinfo{author}{\bibfnamefont{H.}~\bibnamefont{Lamba}},
  \bibinfo{note}{preprint, submitted to J. de Physique}.

\bibitem[{\citenamefont{Diaconis and Freedman}(1999)}]{df}
\bibinfo{author}{\bibfnamefont{P.}~\bibnamefont{Diaconis}} \bibnamefont{and}
  \bibinfo{author}{\bibfnamefont{D.}~\bibnamefont{Freedman}},
  \bibinfo{journal}{SAM Review} \textbf{\bibinfo{volume}{41}},
  \bibinfo{pages}{45} (\bibinfo{year}{1999}).

\end{thebibliography}
\end{document}